\documentclass[conference,onecolumn, twocolumn]{IEEEtran}

\IEEEoverridecommandlockouts
\usepackage{cite}
\usepackage{dirtytalk}
\usepackage{amsmath,amssymb,amsfonts}
\usepackage{algorithmic}
\usepackage{booktabs}
\usepackage{adjustbox}
\usepackage{graphicx}
\usepackage{textcomp}
\usepackage{xcolor}
\def\BibTeX{{\rm B\kern-.05em{\sc i\kern-.025em b}\kern-.08em
    T\kern-.1667em\lower.7ex\hbox{E}\kern-.125emX}}

\begin{document}

\title{LLM-TAKE: Theme-Aware Keyword Extraction Using Large Language Models}

%

\author{
  Reza Yousefi Maragheh*, Chenhao Fang*, Charan Chand Irugu*, \\Parth Parikh, Jason Cho, Jianpeng Xu,  Saranyan Sukumar, \\Malay Patel, Evren Korpeoglu, Sushant Kumar, Kannan Achan\thanks{*All three authors contributed equally to this research. 979-8-3503-2445-7/23/\$31.00 ©2023 IEEE}  \\
  Walmart Global Tech\\
  680 W California Ave\\
  Sunnyvale, CA 94086 \\
  \texttt{\{Reza.Yousefimaragheh,Chenhao.Fang,Charanchand.Irugu,}\\
\texttt{Parth.Parikh, Jason.Cho, Jianpeng.Xu, Saranyan.Sukumar,}\\
\texttt{MPatel, EKorpeoglu, Sushant.Kumar, Kannan.Achan\}@walmart.com}
  }

\maketitle
\IEEEoverridecommandlockouts
\IEEEpubid{\makebox[\columnwidth]{} \hspace{\columnsep}\makebox[\columnwidth]{ }}

\begin{abstract}
Keyword extraction is one of the core tasks in natural language processing. Classic extraction models are notorious for having a short attention span which make it hard for them to conclude relational connections among the words and sentences that are far from each other. This, in turn, makes their usage prohibitive for generating keywords that are inferred from the context of the whole text. In this paper, we explore using Large Language Models (LLMs) in generating keywords for items that are inferred from the items' textual metadata. Our modeling framework includes several stages to fine grain the results by avoiding outputting keywords that are non-informative or sensitive and reduce hallucinations common in LLM’s. We call our LLM-based framework Theme-Aware Keyword Extraction (LLM-TAKE). We propose two variations of framework for generating extractive and abstractive themes for products in an E-commerce setting. We perform an extensive set of experiments on three real data sets and show that our modeling framework can enhance accuracy-based and diversity-based metrics when compared with benchmark models.
\end{abstract}

\begin{IEEEkeywords}
Large Language Models, keyword extraction, context-aware 
\end{IEEEkeywords}

\section{Introduction}
Keyword extraction is defined as the task generating a set of relevant keywords to summarize and characterize an input text\cite{firoozeh2020keyword}. The generated keywords are either extractive or abstractive. Extractive keywords focus on summarizing an input text using the words which exist in the text, while abstractive keywords can be inferred from the text but not exactly contained in the text \cite{over2001duc}. The purpose of keyword extraction is to improve the efficiency of understanding texture information for a human by providing a limited number of keywords. 
In E-commerce settings, this task may help the customers to quickly learn the characteristics of the product from the extracted keywords, and potentially improve the customers' shopping efficiency and experience. 

Classic algorithms for keyword extraction tasks differ in the paradigm of strategies used to extract keywords. Some models extracting the keywords in a supervised manner while others do this in an unsupervised manner \cite{papagiannopoulou2020review}. Also, generally these models either do the task in one stage or two stages. One-stage models performs the task of keywords recall set generation and keyword ranking simultaneously, while two-stage models first generate the recall set and then selects the top keywords in the importance extraction stage \cite{song2023survey}. 

However, most of these models suffer from being a "narrow expert as they are very domain specific and normally work well on texts similar to their training data due to lack of enough background knowledge \cite{tomokiyo2003language, guo2020survey}. Pre-trained Language Models (PLM) have proposed to alleviate this issue by being trained on a larger corpora of text and mapping the text tokens and segments into embedding spaces. However, due to limitations in training data, it is difficult for some PLMs to reason from input text and output results that understand the \textbf{theme} of the entire input text.  

In addition to this, the traditional language models, as well as some PLMs, are notorious for having  short attention spans, which makes it hard for them to draw relational conclusions for words which are apart from each other in a text and thus, make their usage prohibitive in extracting keywords that are obtain from the entire text \cite{daniluk2017frustratingly}. In other words, their usage can be limited when the language task is context dependent \cite{dai2019transformer}.

Large language models have recently shown the potential for language tasks. They outperform the previous PLMs by having the ability of reasoning and wider knowledge background obtained from very large corpora of training data. Their larger architectures allow them to have a larger attention span and can understand the context of the input text better \cite{lampinen2022can}. In particular, they demonstrated beating state-of-the-art models in other language tasks such as summarization and sentiment analysis \cite{touvron2023llama, maragheh2023llm, zhang2023benchmarking}.

In this paper, we propose a multi-stage framework which utilizes the power of the large language models to derive theme-aware keywords for items in E-commerce settings to further help the customers in their shopping journey. We call this LLM-based proposed framework \textit{Theme-Aware Keyword Extraction model} (LLM-TAKE). Our experiments on both proprietary and public data sets show the efficiency of the proposed approach in improving relevant metrics when compared with the state-of-the art models. We discuss how each stage of the framework helps improve the quality of the output keywords and reduce hallucinations.  

The rest of the paper is organized as following. In Section \ref{lit_rev}, we review the related work from the literature. In section \ref{model_fram}, we introduce and discuss the LLM-TAKE framework. In section \ref{emp_res}, we present and analyze the experimental results. Finally, we conclude the paper in section \ref{condl}

\section{Literature Review}\label{lit_rev}
Traditional keyword extraction models are based on statistical or graph-based approaches to the problem. Statistical models rely on various statistical features, such as word frequency, N-grams, location, and document grammar \cite{papagiannopoulou2020review}. However, these features may not adequately capture the complex intricate relationships between words in a document. The fundamental principle of statistical approaches is to determine a given term's score using diverse statistics either within a single document or across several documents. Upon calculating the scores, the method ranks terms according to their scores and highlights the top n terms as essential keywords, with different approaches utilizing distinct methods for calculating N-gram scores (see \cite{ramos2003using}, \cite{campos2020yake}, \cite{li2018efficient}). 
 

In parallel to statistical-based approaches, graph-based keyword extraction has emerged as one of the most effective and widely adopted unsupervised keyword extraction models. Graph-based models represent human language as intricate networks and leverage graphs to encapsulate the multifaceted relationships that exist between words or phrases within a document \cite{cancho2001small}. For instance, \cite{mihalcea2004textrank}, drawing inspiration from PageRank \cite{334}, creates TextRank, which models a document as a graph where nodes symbolize words or phrases and edges represent their connections. \cite{florescu2017positionrank} propose PositionRank, a model that integrates the positional information of a word's occurrences into a biased TextRank, significantly enhancing its performance for longer documents. Following this, several strategies are proposed to enrich the information contained within document graphs. For example, \cite{bougouin2013topicrank} introduces TopicRank, a model aimed at allocating importance scores to topics through candidate keyword clustering. This model applies the TextRank ranking algorithm to evaluate topics and extract keywords by selecting the most indicative candidate from the highest-ranked topics. 

Utilizing the background information obtained from external textual corpora in traditional keyword extraction models has been a long-standing challenge \cite{aizawa2003information} as incorporating external knowledge or additional features lead to further enhancement in keyword extraction task according to \cite{peters2019tune}. Pre-trained embedding models possess a large amount of information, enabling them to accurately represent the relationships between words or phrases. Consequently, pre-trained language model (PLM)-based keyword extraction has witnessed significant progress in recent years according to \cite{papagiannopoulou2020review}.

\cite{mahata2018key2vec} introduce Key2Vec, a method for training embeddings, which proved effective in creating thematic representations of scientific articles and assigning thematic weights to potential keywords. Additionally, they incorporate theme-weighted PageRank \cite{334} to rank these keywords within their framework. However, Key2Vec is only applicable to the domain of scientific articles. \cite{bennani2018simple} presente EmbedRank, which utilizes the cosine similarity between candidate keywords embeddings and the document's sentence embeddings. In light of this development, \cite{sun2020sifrank} proposes an embedding-based model called SIFRank which fuses the sentence embedding model SIF by \cite{arora2017simple} with the autoregressive pre-trained language model ELMo \cite{peters2018deep}, thereby achieving remarkable performance in keyword extraction, particularly for concise documents. \cite{sun2020sifrank} further enhance SIFRank by employing document segmentation and contextual word embeddings alignment, ensuring both speed and accuracy are maintained. In addition, they introduce SIFRank+ for long documents, incorporating position-biased weight to significantly improve its performance on extended texts. 

To make the keywords extraction model more context-aware, AttentionRank by \cite{ding2021attentionrank} employs a pretrained language model to determine the self-attention of a candidate within a sentence's context and cross-attention between a candidate and sentences in a document, thereby assessing the local and global significance of candidates. KeyBERT by \cite{grootendorst2020keybert}, a toolkit for keyword extraction using BERT, first extracts document embedding to obtain a document-level representation. Subsequently, it extracts word embeddings for N-gram words or phrases. KeyBERT then applies cosine similarity to identify the words or phrases most similar to the document, which are considered the best keywords of the entire document. As an upgrade to KeyBERT, \cite{priyanshu2022adaptkeybert} introduce AdaptKeyBERT, a pipeline for training keyword extractors with LLM bases, by integrating regularized attention into a pre-training phase for downstream domain adaptation.

\section{Modeling Framework}\label{model_fram}
In this section, we discuss our LLM-TAKE's framework for generating theme-aware keywords. We go through different stages of the framework we utilized for reducing the hallucinations which is typical to LLM-based methods. 

\subsection{Theme Recall Set Generation}
We start with generating a recall set of candidates of keywords. We use a sophisticated LLM like ChatGPT from \cite{chatgpt} for generating the recall set. As discussed in the introduction we aim at generating two sets of keywords: (i) Abstractive keywords, and (ii) Extractive keywords. We use different prompt strategies for generating each set of keywords. The following figure \ref{ave_prompt_fig} illustrates the prompts we used for each method. 

\begin{figure}[h]
\centering
\includegraphics[width=0.45\textwidth]{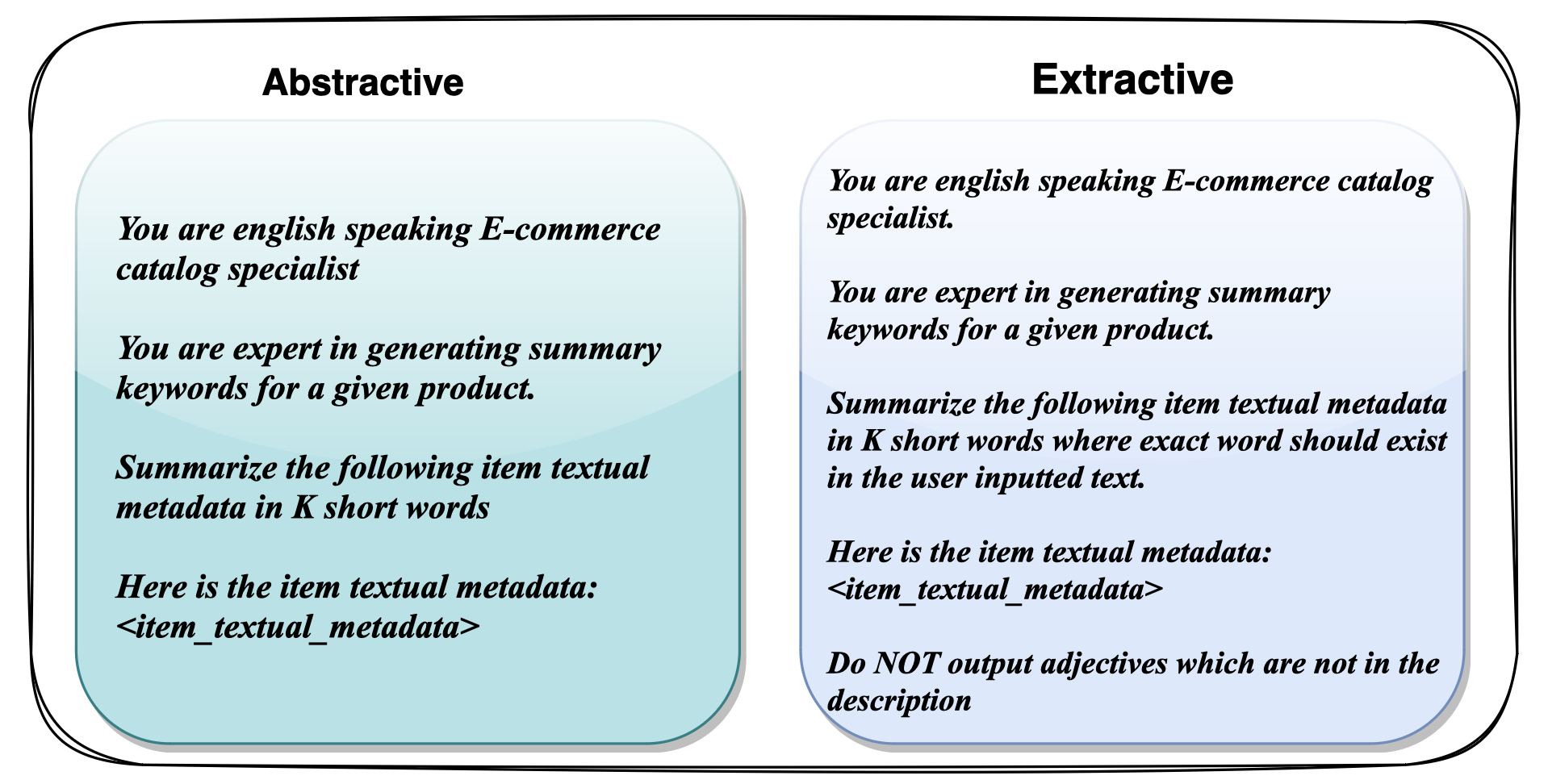}
\caption{Prompts for abstractive method and extractive method.}
\label{ave_prompt_fig}
\end{figure}

Note that generating abstractive keywords may involve more reasoning as the LLM may infer the theme of the product which is not obtained from the inputted text. However, this in turn, may cause more hallucinations, and the generated keywords may not be inferred from text. In this case, extractive keywords may be safer for some of the use cases in industrial applications. 

In our prompting strategies, we use the same base prompts for generating extractive and abstractive keywords, however, we explicitly add the constraint of being present in text when prompting for generation of extractive keywords. Interestingly, we observe that even when explicitly instruct the LLM to generate the output from the text, it sometimes generates the keywords which do not appear in the inputted text. Because of this one may check and drop all the keywords which do not exist in the inputted text even when prompting to obtain the extractive keywords.

\subsection{Hallucination Reduction and Theme Quality Improvement Steps}
LLM-TAKE also includes a series of steps to reduce the hallucinations and improve the quality of final generated keywords as discussed in the following subsections. 

\subsubsection{Constructing the Reference Set of Themes}\label{reference_set}
To alleviate the hallucinations, we first generate a set of keywords with a computationally cheaper model for larger set of products. This computationally cheaper model can be another LLM with lower number of parameters. For reference, we call this computationally cheaper model \say{LLM2}. In our experiments of propitiatory data set, we do this for about 10 million items obtained from similar product categories. This set of item-theme pairs act as reference for final set of items of interest. 

When generating set of theme-aware keywords using main LLM for any given item in the target set of items, we cross check the frequency of the generated themes in the item-theme pair dictionary. If the number of times that the a given theme appears in the reference dictionary is less than a threshold, we eliminate that generated theme. In this way, we avoid themes which are too unique to a given item of interest and hence avoiding too innovative outcomes which have a higher chance being the result of hallucinations. We speculate that some of the unique themes which are generated as the result of hallucinations has a lower chance of reappearing many times in the reference item-theme set. Thus, if a theme has appeared for many items it may be due to prevalence of theme and hence not generated as a result of hallucinations.  

\subsubsection{Eliminating Non-Informative and General Themes}
The objective of the framework is to generate themes that are informative enough and can help the users in their decision journey. Thus, a very general theme may not have the differentiating power to help in user's decision making process. Because of this we eliminate the very general words like \say{Perfect} or \say{Great} which do not add any information value to compare and contrast the product. 

The initial set of general themes is obtained from Top 500 adjectives used in Oxford \cite{dictionary1989oxford}. More keywords are added to this set of general words through manual investigation of sample keyword sets. This constitutes the first block-list we use to eliminate some of the generated keywords by LLM. 

\subsubsection{Eliminating Sensitive Words}
Although many precautions were taken to avoid sensitive responses when training the LLM, we still observed the output of some words that might be interpreted as sensitive in some contexts. To eliminate most of these words, we generate a set of sensitive words using google-profanity-word Github repository from \cite{googleprofanityword}, which contains a full list of bad words and top swear words banned by Google Inc. . This gives us a second set of block-list of words that are used to eliminate some of the generated themes/keywords by LLM.  

\subsection{Theme Importance Extraction}
To revalidate the relevancy of the generated set of themes, and after cross checking the them with the reference item-theme pairs and eliminating the non-informative and sensitive keywords, we obtain the relevancy of generated themes to the item of interest. For doing this, we perform another round of prompts asking the LLM to output a confidence score that measures how descriptive the generated keywords are for the input product. The following figure \ref{imp_extraction} illustrates this step of our framework: 

\begin{figure}[t]
\centering
\includegraphics[width=0.3\textwidth]{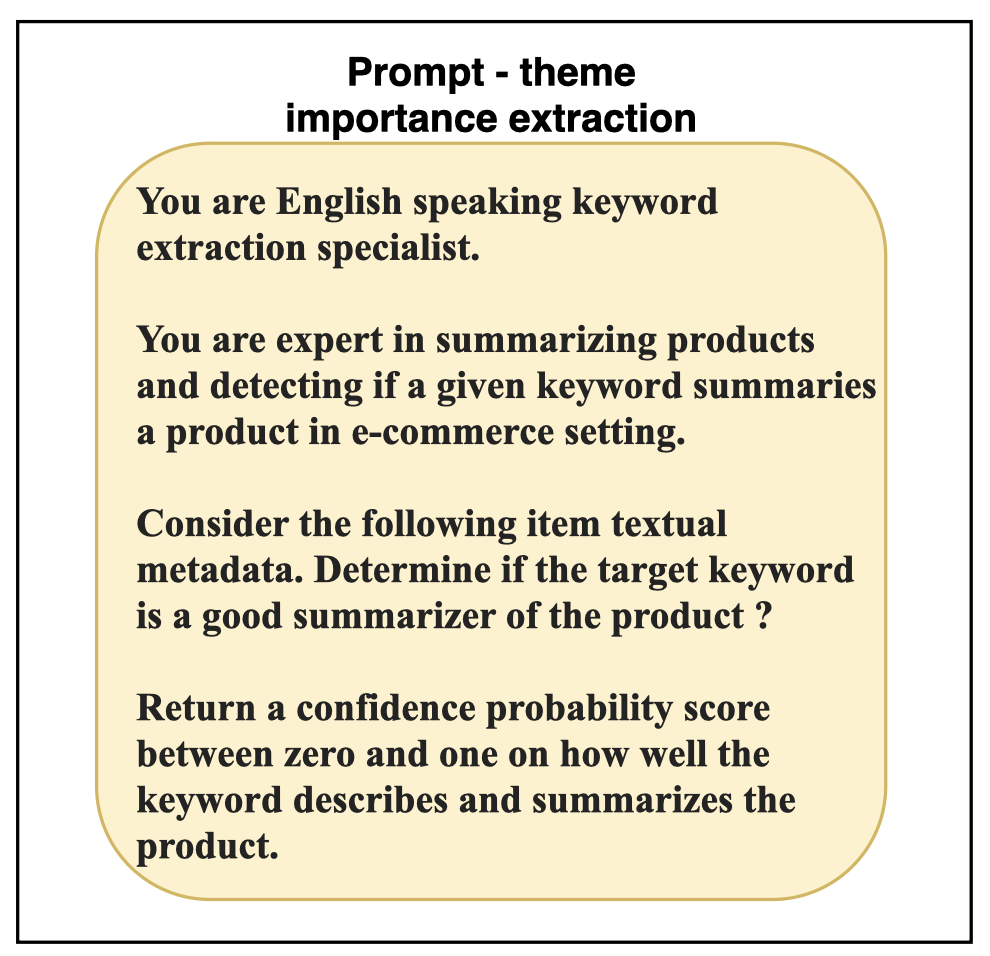}
\caption{Prompt for theme importance extraction.}
\label{imp_extraction}
\end{figure}

\begin{figure}[ht]
\centering
\includegraphics[width=0.4\textwidth]{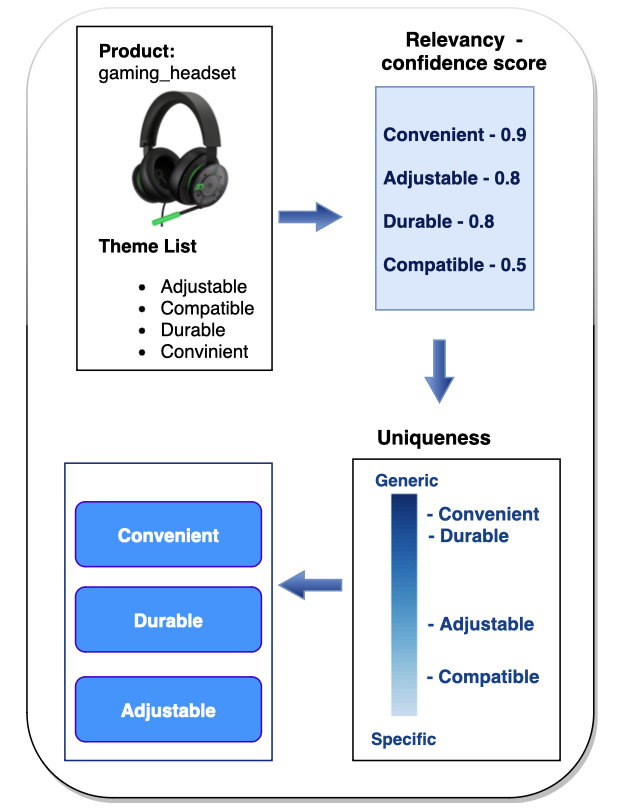}
\caption{Theme Ranking in our framework is based on retrieved score and frequency metric for themes. }
\label{fig_rank}
\end{figure}

Interestingly, when doing this step, we observe a set of generated scores with a very low confidence scores for the relevant products. After doing this step in our proprietary data set, about $10\%$ of item-theme pairs dropped for a score threshold of $0.2$. 

\subsection{Theme Ranking}
We use the generated scores from the previous step, as the primary criteria of ranking. In this case, we may encounter a tie in scores for two themes. The tie is broken by ranking the theme with higher frequency in our reference item-theme set (see subsection \ref{reference_set}) in higher position. In this case, we make sure to de-prioritize more unique items as they may have higher chances of being result of hallucination. If the frequency count is also equal, the keywords are ranked randomly (this is a very rare occurrence according to our experiments as the reference set of item-theme pairs are constructed on a very large number of items).

Note that regardless of rank of the theme, in our framework, we eliminate all of the item-theme pairs that have a score lower than a pre-specified threshold.

\subsection{Keyword Diversification}
Finally, in order to further improve the final set of keyword themes generated for the items, we perform a synonymity check to avoid extracting keywords that are semantically similar. For instance, words 'fun' and 'funny' may appear as a theme for a given product. By doing this step, we eliminate the lower-rank theme which is semantically similar to a higher rank. For obtaining the similarity scores for pairs of words, we use en\_core\_web\_md embedding model from SpaCy Python library to generate words embeddings \cite{Honnibal_spaCy_Industrial-strength_Natural_2020} . 

\begin{figure*}[ht]
\centering
\includegraphics[width=1\textwidth]{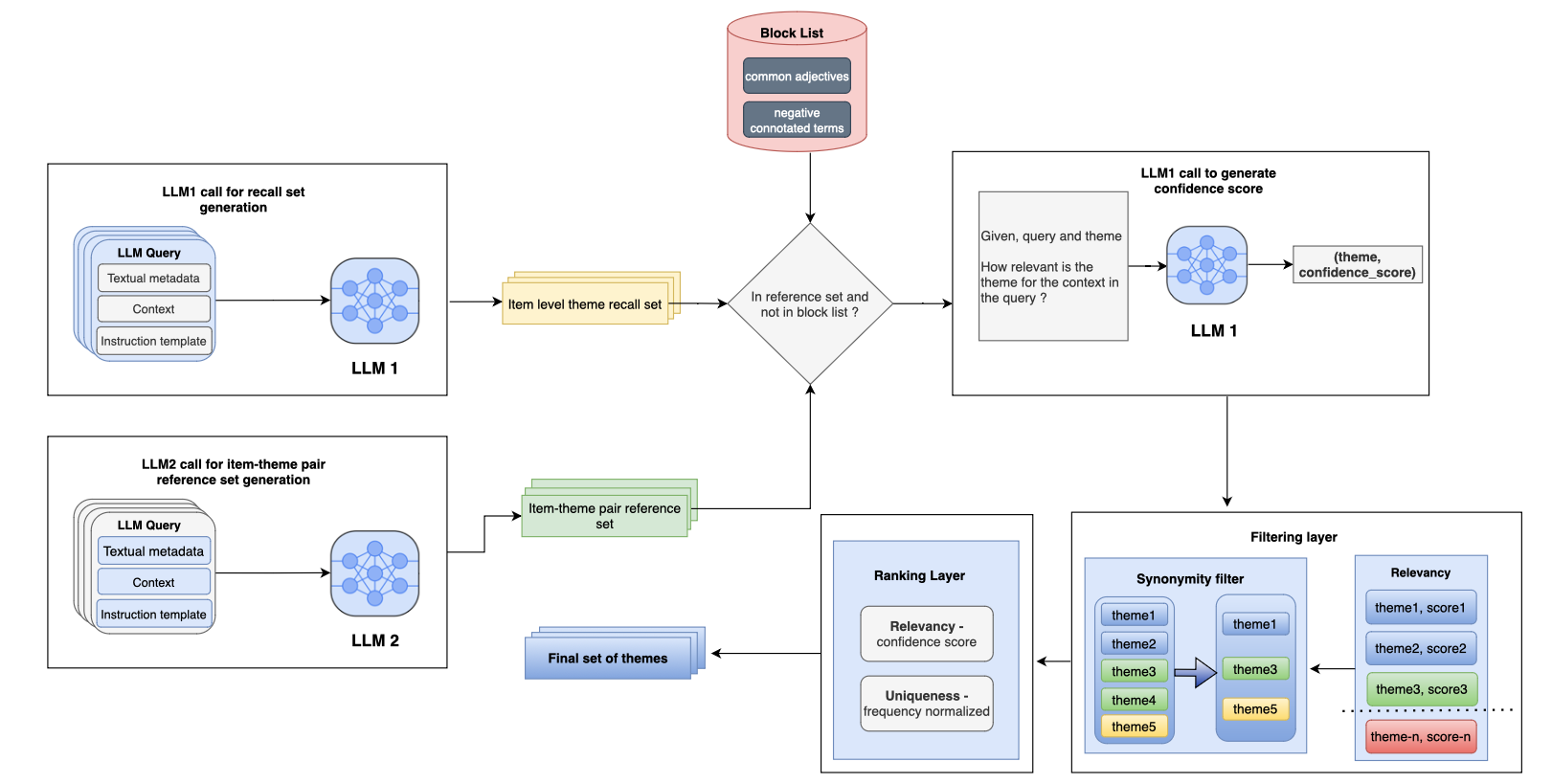}
\caption{An overview of LLM-TAKE Framework}
\label{model_fig}
\end{figure*}

Figure \ref{model_fig} summarizes the workflow of our proposed framework. 

\begin{figure}[ht]
\centering
\includegraphics[width=0.4\textwidth]{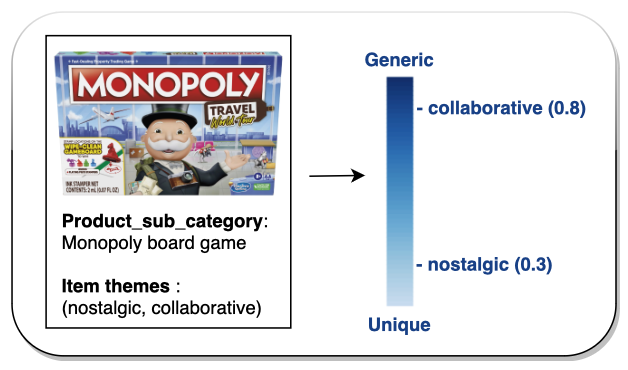}
\caption{By having a reference item-theme pair set we can measure the generality versus uniqueness for a given theme.}
\label{fig_gen}
\end{figure}

\subsection{Approximating Theme Generality vs Uniqueness}
Normally, in E-commerce settings item spaces are divided into broad product categories like Electronics, Toys and Games, Home and Garden etc. These broad product categories are also in turn divided into subcategories. For instance, electronics category can be divided into laptops, headsets etc.    

In this subsection, we want to emphasise on one other capability that the reference item-theme pair set provides us. For all the items in a given product sub categories we can count the frequency of each generated theme for the items of that subcategory. This will give us a measure of the generality versus uniqueness of that generated theme among the items of that product sub category. For instance, consider the items of the \say{board games} sub category. We observe that the theme \say{collaborative} arises more frequently than the theme \say{nostalgic}. Meaning that more number of board games are collaborative and less number of board games are nostalgic (probably have theme of nostalgia in their design or they remind the old childhood times for their players). In this example, we may conclude that \say{collaborative} is a more general theme than \say{nostalgic}. This will enable us to chose different themes when needing to emphasize on more unique or more general themes of the items. See Figure \ref{fig_gen}.  

\section{Empirical Results}\label{emp_res}
In this section, we evaluate the performance of the proposed LLM-TAKE method on real-world data on two proprietary data sets obtained from an E-commerce platform, and DUC2001 public data. One of the proprietary includes set of items from Electronics Category (denoted by P-Electronics) and the other includes items from Toys and Games category (denoted by P-Toys). The performance of LLM-TAKE is compared against various state-of-the-art models keyword extraction models. The number of documents of the these three datasets and the number of unique keywords generated by LLM-Take and all the benchmark models are shown in table ~\ref{dataset-stats}.

\subsection{Datasets}

The proprietary dataset is comprised of 188 high-traffic items from an E-commerece platform, with their respective textual meta-data serving as documents for keyword extraction. This dataset includes 94 items from the Toys category of items and another 94 items from the Electronics category. 

DUC2001 \cite{wan2008single} public dataset is also used to evaluate the performance of different models. This dataset is manually annotated in the ExpandRank paper \cite{wan2008single}. Originally designed for document summarization, the DUC2001 dataset consists of 308 news articles covers 30 distinct news topics. On average, each article contains 740 words.
\begin{table}[ht]
  \caption{Number of documents, number of unique keywords generated for each data set, and the average number of generated keywords for each dataset}
  \label{dataset-stats}
  \centering
  \begin{tabular}{cccc}
    \toprule
    Dataset  &  \# Documents   &  \# Unique Keywords    &  \# of keywords  \\
      &  &      &  per document  \\
    \midrule
    \textbf{P-Electronics}&94& 179  & 3.05     \\
    \textbf{P-Toys}  &94   & 184 & 3.24      \\
    \textbf{DUC2001} &308    & 2488      & 8.08  \\
    \bottomrule
\end{tabular}
\end{table}
\subsection{Benchmark Models}
We compare our model with both traditional keyword extraction models and embedding based keyword extraction models. We evaluate YAKE \cite{campos2020yake}, SIFRank, SIFRank+\cite{sun2020sifrank}, KeyBERT \cite{grootendorst2020keybert} and AdaptKeyBERT \cite{priyanshu2022adaptkeybert}. For YAKE, we set the window size to be 1, deduplication threshold equal to 0.9 and n-gram length equal to 1. For SIFRank and SIFRank+, we use the codebase and the same parameters suggested by the authors. We only change the number of returned Top-n words to 3. For KeyBERT, we use the toolkit developed by \cite{grootendorst2020keybert}. And we implement the AdaptKeyBERT using an open-source python library AdaptKeyBERT by \cite{priyanshu2022adaptkeybert}, which was built upon KeyBERT toolkit\cite{grootendorst2020keybert}. We implement the LLM-TAKE method with GPT-3.5 API by OpenAI \cite{chatgpt}. We also generate the reference set for item-theme pairs using a computationally cheaper LLM. We test both abstractive and extractive LLM-TAKE methods. Since the quantity of keywords annotated varies across documents, the number of keywords extracted (N) is set to 3. In this study, we utilize macro Precision (P), Recall (R), and F1 score (F1) at 3 to evaluate the models \cite{derczynski2016complementarity}.

\subsection{Performance Comparison}

A team of product experts manually evaluate and select the top keywords for each item for the proprietary dataset. 322 unique keywords are selected by this team across all 188 items, averaging 3.14 keywords per document. For the 94 Toy items, 186 distinct keywords are selected by product experts, resulting in an average of 3.24 keywords per document. In addition, the 94 electronics items have 179 unique keywords selected, with an average of 3.05 keywords per document.

\begin{table*}[t]
  \caption{Comparison of LLM-TAKE with Other Benchmark Models}
  \label{performance-table}
  \centering
  \begin{tabular}{cccccccccc}
    \toprule
    \multicolumn{1}{c}{} & \multicolumn{3}{c}{P-Electronics}  & \multicolumn{3}{c}{P-Toys}  & \multicolumn{3}{c}{DUC2001}\\
    \cmidrule{2-4}       \cmidrule{5-7}       \cmidrule{8-10}       
    Method  & P &  R   &  F1 & P &  R   &  F1 & P &  R   &  F1  \\
    \midrule
    YAKE    & 0.142&0.139& 0.07   & 0.158&0.147& 0.076   & 0.264&0.122& 0.083    \\
    KeyBERT    & 0.129&0.13& 0.065    & 0.128&0.117& 0.061    & 0.257&0.121& 0.082    \\
    AdaptKeyBERT   & 0.078&0.083& 0.04    & 0.11&0.098& 0.052    & 0.262&0.124& 0.084    \\
    SIFRank     & 0.119&0.105& 0.056   & 0.188&0.166& 0.088    & 0.131&0.06& 0.041    \\
    SIFRank+    & 0.119&0.105& 0.056   & 0.188&0.166& 0.088    & 0.131&0.06& 0.041    \\
    
    \midrule
    
    LLM-TAKE extractive     & 0.371&0.274& 0.157    & 0.344&0.239& 0.141    & \textbf{0.305}&\textbf{0.15}& \textbf{0.1 }   \\
    LLM-TAKE abstractive   & \textbf{0.504}&\textbf{0.464}& \textbf{0.241}    & \textbf{0.496}&\textbf{0.42}& \textbf{0.227}     & 0.079&0.037& 0.025    \\
    \bottomrule
  \end{tabular}
\end{table*}

Table~\ref{performance-table} presents the top 3 Precision, Recall, and F1 scores for all extraction methods discussed in the paper. The table is divided into two sections: the first displays benchmark results, while the second demonstrates LLM-TAKE's performance in both extractive and abstractive forms. Among the benchmark models, we observe that the YAKE model, despite being easy to implement obtains the comparable scores to other benchmark models. The SIFRank and SIFRank+ display identical outcomes, as we only truncate the top 3 keywords, which are ranked highly in both models.

In the DUC2001 dataset, the AdaptKeyBERT maintains a slight edge over other benchmark models. The abstractive version of our LLM-TAKE model has achieved state-of-the-art results in P-Electronics and P-Toys. Meanwhile, the extractive version of LLM-TAKE secures the best performance in the DUC2001 dataset, which consists of long news documents. The distinct behavior of abstractive and extractive LLM-TAKE methods can be attributed to the differences in the dataset annotation process. In the DUC2001 dataset, two graduate students \cite{wan2008single} employed extractive methods (not abstractive) for annotating the data. 

\begin{figure}[ht]
\centering
\includegraphics[width=0.25\textwidth]{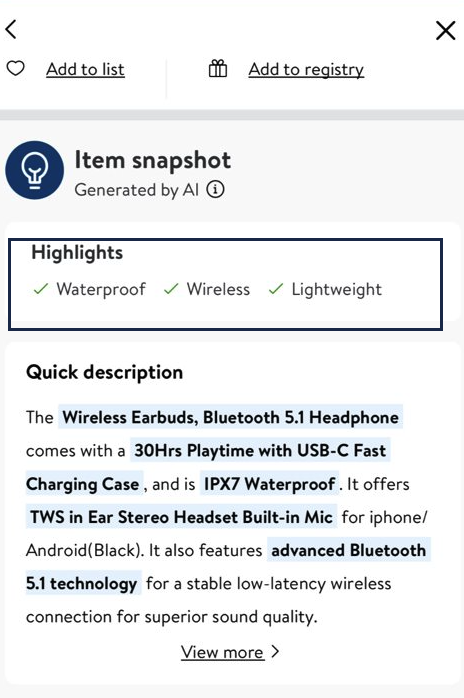}
\caption{Screen shots of generated keywords for a TV product from the E-commerce platform.}
\label{fig_gen1}
\end{figure}

On the other hand, for P-Electronics and P-Toys, the labeling process involved combining outputs from all baseline models, as well as LLM-TAKE abstractive and extractive outputs. This combined output list was then shuffled and forwarded to a team of product experts without informing them about the source method for each keyword. A team of product experts subsequently selected and ranked the top three results to serve as labels.

\subsection{Online Experiment}
After the offline experiments, which show the state-of-the-art performance, we decide to advance to an online experiment. In these online tests, we display the top three keywords derived from product descriptions as highlights for specific items, see Figure \ref{fig_gen1}. We introduce this feature across over 200 product categories on the e-commerce platform.

Table~\ref{online-experiments} shows the outcomes of the online experiment, which reveals a statistically significant enhancement in various aspects, including different business metrics. We mask the name of business metrics for proprietary reasons. These findings serve as a strong indication that the LLM-TAKE methodology is indeed highly effective. This further emphasizes the importance and potential impact of the LLM-TAKE in the realm of e-commerce applications.

\begin{table}[h]
  \caption{Online experiment results}
  \label{online-experiments}
  \centering
  \begin{tabular}{ccc}
    \toprule            
    Metrics  & Percentage Lift & P-Value  \\
    \midrule
    Business Metric 1 &9.76\%  &0.011  \\
    Business Metric 2  &6.99\%  & 0.025   \\
    Business Metric 3 &5.17\%  & 0.047\\
    \bottomrule
  \end{tabular}
\end{table}

\section{Conclusion}\label{condl}
Classic keyword extraction model suffer both from limited training data as well as short attention span. Because of this, their usage becomes limited when it comes to generating context-aware and theme-aware keywords from any text document. In this paper, we propose a LLM-based framework for generating context-based and theme-aware keywords of products in E-commerce settings. We call our modeling framework Theme-Aware Keyword Extraction (LLM-TAKE) method. We propose two variations for  generating extractive and abstractive keywords from the textual meta-data of products. We discuss different stages of the framework implemented to reduce the chance of generating keywords that are sensitive or not informative. We also discuss our methodology to reduce hallucinations by cross-checking any given LLM-generated themes a reference set of item-theme pairs which are generated by computationally cheaper model. Furthermore, we illustrate how this reference item-theme pair can be used as a measure of generality versus uniqueness for any generated theme. 

Our experiments on three annotated real world data sets show that our framework lead to higher accuracy based metrics and thus, proves the capability of large language models in generating theme-aware keywords for some data sets.

\section{Limitations}
Common risk associated with LLMs is a possibility of hallucination where generated text can be factually inaccurate or erroneous in nature. Though we have incorporated multiple steps in our methodology to counter the hallucination which includes maintaining a reference set, eliminating non-informative words, and de-prioritizing more unique words as explained in detail in the earlier sections, probability of hallucination would still be non-zero as LLM might be challenged by limited contextual understanding due to the noise in the input.

\bibliography{./IEEEabrv,./IEEEexample}
\end{document}